\documentclass{iau-JDSS}
\usepackage{graphicx}

\title[Probing fundamental physics with pulsars] %% give here short title %%
{Probing fundamental physics with pulsars}

\author[Lorimer \& McLaughlin]   %% give here short author list %%
{Duncan R. Lorimer$^1$ \and Maura A. McLaughlin$^1$}

\affiliation{$^1$Department of Physics, 210 Hodges Hall, Morgantown,
WV 26506, USA\break email: duncan.lorimer@mail.wvu.edu and 
maura.mclaughlin@mail.wvu.edu
}

\pubyear{2009}
\volume{Volume 15}  %% insert here IAU Highlights of Astronomy Volume Number
\pagerange{119--126}
\date{?? and in revised form ??}
\jname{Highlights of Astronomy, Volume 14}
\editors{Ian F Corbett, ed.}
\begin{document}

\maketitle

\begin{abstract}
Pulsars provide a wealth of
information about General Relativity, the equation of state of superdense matter,
relativistic particle acceleration in high magnetic fields, the
Galaxy's interstellar medium and magnetic field, stellar and binary
evolution, celestial mechanics, planetary physics and even cosmology.
The wide variety of physical applications currently being investigated
through studies of radio pulsars rely on: (i) finding
interesting objects to study via large-scale and targeted surveys;
(ii) high-precision timing measurements which exploit their remarkable
clock-like stability. We review current surveys and
the principles of pulsar timing and highlight progress made in the
rotating radio transients, intermittent pulsars, tests of relativity,
understanding pulsar evolution, measuring neutron star masses and the
pulsar timing array.
\end{abstract}

\firstsection % if your document starts with a section,

\section{Pulsar surveys}

A wide variety of successful
pulsar surveys are being carried out using most of the major
radio telescopes. We summarize the current status and 
future prospects here.

\begin{itemize}

\item[\bf Parkes] The 1.4-GHz multibeam receiver
 has discovered over 800 pulsars
in the past decade (see Lyne 2008).
Reanalyses have resulted in a new class
of pulsars (McLaughlin
et al.\ 2006; see \S \ref{sec:rrats}) and continue to find new pulsars
using new techniques (Keith et al.\ 2009). New
data acquisition systems provide significantly
improved sensitivity to millisecond pulsars (MSPs) and are being used to 
resurvey the sky. 

\item[\bf Effelsberg] 
A similar data acquisition system is now being 
installed on the seven-beam receiver to provide a complementary
survey of the Northern sky.

\item[\bf Arecibo] Since 2004,
 Arecibo has been surveying the Galactic plane with a 1400~MHz seven-beam receiver, and
 currently 50 pulsars have been discovered
 (Cordes et al.\ 2006).  Among these new
discoveries is a double neutron star binary (Lorimer et al.\ 2006) an
eccentric binary MSP (Champion et al.\ 2008) and a
number of RRATs~(Deneva et al.\ 2009). Several hundred pulsars are
expected from this survey over the next five years.

\item[\bf Green Bank] A 350~MHz receiver  
 has enabled surveys of the Northern
Galactic plane (Hessels et al.\ 2008) and a drift scan survey of over
10,000 square degrees (Boyles et al.\ 2008), resulting in a combined
 60 new pulsars so far, including five MSPs.
The number of pulsars found in
these surveys should at least double in the next two
years.

\item[\bf Giant Metrewave Radio Telescope] Surveys  at 610~MHz at low
  (Joshi et al.\ 2009) and
 intermediate  (Bhattacharya, private communication) latitudes 
have been successful, but are plagued with severe radio-frequency interference (RFI).
A new survey 
 at 327~MHz will cover 1600 square degrees in a better RFI environment, and
 is expected to detect roughly 250 pulsars and 30 RRATs (McLaughlin, private
communication).

\end{itemize}

\section{Principles of pulsar timing}

Once a new pulsar is found, it is observed at least once or twice per month
over the course of a year. During
each observation, pulses from the neutron star traverse the
interstellar medium before being received at the radio telescope, where
they are dedispersed and added in phase to form a mean pulse profile. The
time-of-arrival (TOA) is defined as the time of some fiducial
point on the integrated profile.
 Since the profile has a stable form
at any given observing frequency, the TOA can be accurately determined
by cross-correlation of the observed profile with a 
``template'' profile obtained from the addition of profiles from many observations
at a particular observing frequency.

The TOAs are
first corrected to the solar system
barycenter.
 Following the accumulation of a number of TOAs, a
 simple model is usually sufficient to fit the
TOAs during the time span of the observations and to predict the
arrival times of subsequent pulses. The model is a Taylor expansion of
the rotational frequency $\Omega = 2 \pi/P$ about a model value
$\Omega_0$ at some reference epoch ${\cal T}_0$.  Based on this
model, and using initial estimates of the position, dispersion measure
and pulse period, a ``timing residual'' is calculated for each TOA as
the difference between the observed and predicted pulse phases.

Ideally, the residuals should have a zero mean and be
free from systematic trends. To reach this point, the
model needs to be refined in a bootstrap fashion. Early 
residuals show a number of trends indicating an
error in one or more of the parameters, or a parameter not
yet added to the model. For further details, see
 Lorimer \& Kramer (2005).

\section{Highlights from the past few years}

\subsection{Rotating radio transients} \label{sec:rrats}

The discovery of the rotating radio transients (RRATs) in a reanalysis
of the Parkes multibeam survey data (McLaughlin et al.\ 2006)
demonstrates the wealth of new sources awaiting
detection. 
 The radio emission from RRATs
is typically only visible for $<1$~s per day making them extremely
difficult to study.  Their detection was made possible by searching
for dispersed radio bursts (Cordes \& McLaughlin 2003),
which often do not show up in conventional Fourier-transform based
searches (Lorimer \& Kramer 2005).

Since the initial discovery, a significant effort has gone in to
searching for and characterizing more RRATs.
 Nearly 30 are
known
(Hessels et al.\ 2008, Deneva et al.\ 2009, Keane et al.\ 2009) but
only seven have  timing
solutions, with four of these only recently achieved (McLaughlin et al.\ 2009).
 It is clear  
(Figure~1a), that the RRATs exhibit varied spin-down properties. Recently,
Lyne et al.~(2009) reported the detection of two glitches
in RRAT J1819$-$1458. While these events are similar in magnitude to
the glitches seen in young pulsars and magnetars, they are accompanied
by a long-term {\it decrease} in the spin-down rate,
suggesting that
it previously occupied the phase space populated by the magnetars.
Further observations could confirm this
``exhausted magnetar'' hypothesis.

\begin{figure}
\includegraphics[width=\textwidth]{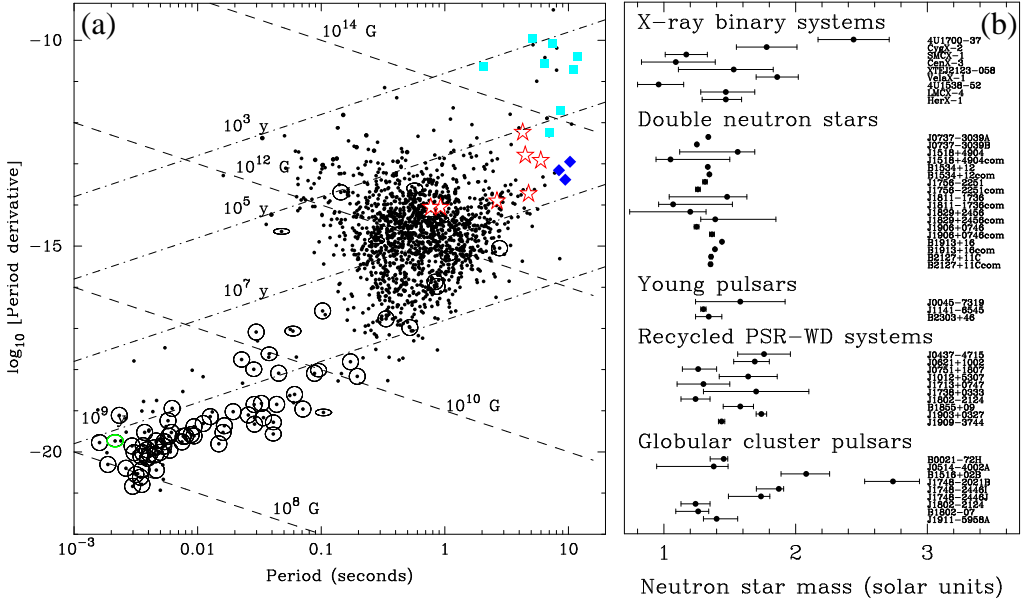}
  \caption{(a) $P-\dot{P}$ diagram of the neutron star
      population, with radio pulsars as dots, RRATs as open red stars, soft gamma-ray repeaters as blue squares and
anomalous X-ray pulsars as cyan triangles.  Binaries are marked by circles, with ellipticity equal to that of the orbit and J1903+0327 marked in green. (b) Distribution of neutron
      star masses as inferred from timing observations of binary
      pulsars and X-ray binary systems.}
\end{figure}

  McLaughlin et
al.~(2009) find that the probabilities that the periods and magnetic fields of
 RRATs and normal pulsars are drawn from the same parent
distributions are small ($<10^{-3}$), with the RRATs having longer
periods and higher magnetic fields. 
This effect appears
to be real and not due to a bias against short period objects.  
The other spin-down derived parameters 
of normal pulsars and RRATs are consistent.

\subsection{Intermittent pulsars}

Another new class of radio pulsars, reviewed by
Kramer at this meeting, are the intermittent pulsars.
The prototype, PSR~B1931+24 (Kramer et al.~2006a), shows a
quasi-periodic on/off cycle in which the spin-down rate increases by
$\sim 50$\% when the pulsar is on.
 A spectral
analysis reveals a persistent periodicity that slowly varies with time
in the range of 30--40 days.  The pulsar 
switches off in less than 10~s, a timescale  too small
for precession and indicative of a relaxation-oscillation of unknown
nature.
The dramatic change in spin-down rate points
to a large increase in the magnetospheric particle outflow when the
pulsar switches on. The
 changes allow us to estimate the current density, which
 coincides with the
 predictions by Goldreich \& Julian (1969) for pulsar
magnetospheres and proves, for the first time, that the pulsar wind
plays a substantial role in spin down.

Since PSR~B1931+24 is only visible for 20\% of the time, we 
 estimate that there should be at least 5 times as many similar
objects.  
 Timing archives should be carefully mined to find pulsars with
similar characteristics. A number of examples have been found,
one of which, PSR~J1832+0029, 
switched off in 2004 after 270 days of consistent 
 detections and switched on again in
2006. Due to this apparent long timescale,
  we have not yet fully sampled the
on/off cycles.  
 RW find the implied magnetospheric
charge density for PSR~J1832+0029 to be almost
four times higher than the Goldreich-Julian density! This cannot be
easily reconciled by any of the current models (the debris disks of
Cordes \& Shannon 2006; the death valley scenario of Zhang et al. 2007
or the accretion model of Rea et al.\ 2008) and highlights the need for
further work.

\subsection{An eccentric millisecond pulsar}

The 2.15-ms pulsar J1903+0327, found in the
Arecibo multibeam survey (Champion et al.\ 2009), is
distinct from all other Galactic MSPs in that its 95-day orbit
has an eccentricity of 0.43! Optical observations show a possible counterpart
which is consistent with a 1~M$_{\odot}$ star.  While similar systems
have been observed in globular clusters,
presumably a result of exchange interactions, the
standard hypothesis in which MSPs are ``recycled''
via the accretion of mass in a low-mass X-ray binary system
cannot account for J1903+0327.
A triple system
 now appears to be ruled out due to the lack of any change
in orbital eccentricity in the system (Gopakumar et al.\
2009).
It is possibile that the binary system 
was produced in an exchange
interaction in a globular cluster and subsequently ejected, or the
cluster has since disrupted. Statistical estimates of
the likelihood of both these channels are roughly 1--10\%.
Finally, as discussed by Liu and Li at this meeting, a scenario
involving accretion from a supernova fall-back disk is also viable.

\subsection{Tests of relativity}

Although most binary pulsars can be
adequately timed using Kepler's laws, there are a number which require
an additional set of ``post-Keplerian'' (PK) parameters which have
a distinct form for a given relativistic theory of
gravity (Damour \& Taylor 1992).
In General Relativity (GR) the PK formalism (see Lorimer \& Kramer
2005) gives the relativistic
advance of periastron, $\dot{\omega}$,
the time dilation and gravitational redshift parameter, $\gamma$,
the rate of orbital decay due to gravitational radiation, $\dot{P_{\rm b}}$,
and the two Shapiro delay parameters, $r$ and $s$.
Some combinations, or all, of the PK parameters have now been 
measured for a number of binary pulsar systems.

Given the precisely measured Keplerian parameters, the only two unknowns
are the masses of the pulsar and its companion,
$m_\mathrm{p}$ and $m_\mathrm{c}$. Hence, from a measurement of just
two PK parameters one can
solve for the two masses. From the Keplerian mass function
%
%\begin{equation}
  %f_\mathrm{mass} = \frac{4\pi^2}{G} \frac{x^3}{P_\mathrm{b}^2} = 
  %\frac{(m_\mathrm{c} \sin i)^3}{(m_\mathrm{p} + m_\mathrm{c})^2},
  %\label{equ:massfn}
%\end{equation}
%
one can then find the orbital inclination angle $i$.
If three (or more) PK parameters are measured, the system is
``overdetermined'' and can be used to test GR (or
 any other
theory of gravity) by comparing
the third PK parameter with the predicted value based on
the masses determined from the other two. 

As discussed by Kramer's contribution in JD06,
currently the best binary pulsar system for
strong-field  GR tests is the double
pulsar J0737$-$3039. In this system, with two independent pulsar
clocks, five PK parameters of the $22
\mbox{-}\mathrm{ms}$ pulsar ``A'' have been measured as well as two
additional constraints from the  mass function and projected
semi-major axis of the $2.7 \mbox{-}\mathrm{s}$ pulsar ``B''. The
measurement of the projected semi-major axes gives a
mass ratio $R=1.071 \pm 0.001$. The mass ratio measurement is unique
to the double pulsar and rests on the assumption that
momentum is conserved. The observations of $\dot{\omega}$ and $R$
yield the masses of A and B as $m_\mathrm{A}=1.3381\pm0.007\,M_\odot$
and $m_\mathrm{B}=1.2489\pm0.0007\,M_\odot$. From these values,
 the expected values of
$\gamma$, $\dot{P}_\mathrm{b}$, $r$ and $s$ may be calculated and
compared with the observed values. These four tests of GR
 all agree with the theory to within the
uncertainties. Currently the tightest constraint is the Shapiro delay
parameter $s$ where the observed value is in agreement with GR at the
0.05\% level (Kramer et al.\ 2006b).

Another unique feature of the double pulsar system is the interaction
between the two pulsars. The signal
from A is eclipsed for 30~s each orbit by the magnetosphere of
B (Lyne et al.\ 2004) and the radio pulses from B are modulated
by the relativistic wind from A during one phases of the
orbit (McLaughlin et al.\ 2004). 
These provide unique insights into plasma
physics.
  By modeling of the change in eclipse
profiles of A over four years, Breton et al.~(2008)
fit a simple model to determine the
precession of B's spin axis about the orbital angular momentum vector.
This measurement agrees, within the 13\% measurement
uncertainty, with to the GR prediction.

\subsection{Massive neutron stars}

Multiple PK parameters measured for a number of binary
pulsars provide precise constraints on neutron star
masses (Thorsett \& Chakrabarty 1999). As shown in Figure~1b 
(from Lorimer 2008), the young
pulsars and the double neutron star binaries are consistent
with, or just below, the canonical $1.4\,M_\odot$, but
the MSPs in binary systems have, on average,
significantly larger masses. 

Several eccentric binary systems
in globular clusters have their masses constrained from
measurements of the relativistic advance of periastron and
the Keplerian mass function. In these cases, the condition $\sin i<1$
sets a lower limit on the companion mass
$m_\mathrm{c}>(f_\mathrm{mass} M^2)^{1/3}$ and an upper limit
on the pulsar mass. Probability density functions for both 
$m_\mathrm{p}$ and $m_\mathrm{c}$ can also be estimated in a statistical sense
by {\it assuming} a random distribution of orbital inclinations.
An example is the eccentric binary MSP in 
M5 (Freire et al.\ 2008)
where the nominal pulsar mass is $2.08 \pm 0.19$~M$_{\odot}$.
If these can be confirmed by the measurement of
other relativistic parameters, these supermassive neutron stars
will have important constraints on the equation of state of superdense
matter.

Currently the largest measurement of a radio pulsar mass is the
eccentric MSP binary J1903+0327~(Champion et
al.\ 2008). Recent timing measurements of the relativistic periastron
advance and Shapiro delay in this system by Freire et al. (2009) yield a  
 mass of $1.67 \pm 0.01$~M$_{\odot}$.  When placed on
the mass--radius diagram for neutron stars~(Lattimer \& Prakash 2007)
this pulsar appears to be
incompatible with at least four equations of state.
 Optical measurements of the companion are required to rule
out classical contributions to $\dot{\omega}$, and further timing
measurements are required to 
verify this result.

\subsection{Pulsar timing and gravitational wave detection}

As discussed by Andersson at this meeting, the direct detection of GWs
is one of the foremost goals of modern  physics.
Many cosmological models predict that the Universe is presently filled
with an ultra low-frequency (nHz) stochastic gravitational wave (GW)
background produced during the big bang era (Peebles 1993). A
significant component (Jaffe \& Backer 2003) is the gravitational
radiation from massive black hole mergers due to Galaxy collisions
at a redshift $z \sim 1$.
Pulsars can be used as natural GW detectors of
this background  (Sazhin 1978;
Detweiler 1979). 
 pulsar acts as a reference clock,
sending out regular signals which are monitored by an observer on the
Earth over some time-scale $T$. Passing
GWs perturb the local spacetime and cause a change
in the observed rotational frequency of the pulsar. For regular pulsar
timing observations with typical TOA uncertainties of
$\epsilon_\mathrm{TOA}$, this detector would be sensitive to GWs
with  amplitudes $h \gtrsim \epsilon_\mathrm{TOA}/T$ and
frequencies $f \sim 1/T$ (Bertotti et al.\ 1983; Blandford et al.\
1984).

A natural extension of this concept is a ``timing array'' of 
 a number of pulsars distributed over the
sky (Hellings \& Downs 1983), allowing
 cross-correlation of the residuals for pairs of
pulsars (Foster \& Backer 1990). 
It should therefore be possible to separate the timing noise of each
pulsar from the common signature of the quadrupolar GW background
 from the effects of clock errors (which have a monopolar
signature) and solar system ephemeris errors (which have dipolar
signature).

As reviewed by Manchester in JD06,
 three main groups are
collaborating to form an international pulsar
timing array. The European Pulsar Timing Array
consists of four radio telescopes which will be combined
to produce a 300-m class telescope. In Australia, the
Parkes Pulsar Timing Array uses the 64-m
Parkes telescope to time Southern pulsars. In North America, the Green
Bank and Arecibo telescopes are used in the NANOGrav collaboration.
The best existing limits (Jenet et al. 2006) 
constrain the merger rate of supermassive black hole binaries at high
redshift, investigate inflationary parameters and place limits on the
tension of currently proposed cosmic string scenarios.

Based on current expectations of the likely strength of the GWB, a
detection requires timing of 20 MSPs with 100~ns
residuals for a period of five years~(Jenet et al.\ 2005). In general,
for residuals $\delta t$, data span $T$, and number of pulsars $N$,
the sensitivity scales roughly as $\delta t^2/(N T^4)$ (Kaspi et
al.~1994).  Of the $\sim$30 MSPs that are regularly
timed internationally, five have achieved a timing precision of 100 ns
or less. With improved algorithms and more sensitive observations,
such precision may soon be achieved for 5--10 more.
A key goal of the ongoing surveys is to find more MSPs
to add to the array. This should be accomplished by the various surveys outlined
in Section~1, and GW detection may be achievable within the next 5 -- 10 years.

\begin{acknowledgments}
We thank the AAS for travel support. Our research is supported by
the National Science Foundation, the Research Corporation for 
Scientific Advancement, West Virginia EPSCoR and the Alfred
P. Sloan foundation.
\end{acknowledgments}

\end{document}